\documentclass[conference]{IEEEtran}
\IEEEoverridecommandlockouts

\usepackage{caption}
\usepackage{mdframed}

\newmdenv[linecolor=black,linewidth=0.5pt]{algorithmframe}
\usepackage{cite}
\usepackage{amsmath,amssymb,amsfonts}
\usepackage{algorithmic}
\usepackage{graphicx}
\usepackage{textcomp}
\usepackage{xcolor}
\def\BibTeX{{\rm B\kern-.05em{\sc i\kern-.025em b}\kern-.08em
    T\kern-.1667em\lower.7ex\hbox{E}\kern-.125emX}}
\begin{document}

\title{AI-enabled Priority and Auction-Based Spectrum Management for 6G
}

\author{\IEEEauthorblockN{Mina Khadem}
\IEEEauthorblockA{\textit{Department of Electrical Engineering} \\
\textit{Tarbiat Modares University}\\
Tehran, Iran \\
0009-0008-1872-699X}
\and
\IEEEauthorblockN{Farshad Zeinali}
\IEEEauthorblockA{\textit{Department of Electrical Engineering} \\
\textit{Tarbiat Modares University}\\
Tehran, Iran \\
0009-0008-6349-9837}
\and
\IEEEauthorblockN{Nader Mokari}
\IEEEauthorblockA{\textit{Department of Electrical Engineering} \\
\textit{Tarbiat Modares University}\\
Tehran, Iran \\
0000-0001-5364-8888}
\and
\IEEEauthorblockN{}
\IEEEauthorblockA{\textit{} \\
\textit{\textcolor{white}{University Technology}}\\
 \\
}
\and
\IEEEauthorblockN{Hamid Saeedi}
\IEEEauthorblockA{\textit{College of Engineering and Technology} \\
	\textit{University of Doha for Science and Technology}\\
	Doha, Qatar \\
	0000-0003-2706-228X}
\and
\IEEEauthorblockN{}
\IEEEauthorblockA{\textit{} \\
	\textit{}\\
	 \\
}
}

\maketitle

\begin{abstract}
In this paper, we present a quality of service (QoS)-aware priority-based spectrum management scheme to guarantee the minimum required bit rate of vertical sector players (VSPs) in the 5G and beyond generation, including the 6th generation (6G). VSPs are considered as spectrum leasers to optimize the overall spectrum efficiency of the network from the perspective of the mobile network operator (MNO) as the spectrum licensee and auctioneer. We exploit a modified Vickrey-Clarke-Groves (VCG) auction mechanism to allocate the spectrum to them where the QoS and the truthfulness of bidders are considered as two important parameters for prioritization of VSPs. The simulation is done with the help of deep deterministic policy gradient (DDPG) as a deep reinforcement learning (DRL)-based algorithm. Simulation results demonstrate that deploying the DDPG algorithm results in significant advantages. In particular, the efficiency of the proposed spectrum management scheme is about $\%85$ compared to the $\%35$ efficiency in traditional auction methods.
\end{abstract}

\section{Introduction}
Recent efforts by regulatory authorities, industry stakeholders, and the scientific community address the imperative challenge of optimizing the utilization of limited spectrum resources. This heightened focus arises in response to the rapid development of wireless devices and the growing demand for mobile broadband services. The traditional radio spectrum management approaches involve its division among various services, which are subsequently designated to specific spectrum bands, categorized as either licensed or unlicensed. Therefore, to effectively cater to the diverse needs of various sectors within the evolving landscape of the $6$th generation network ($6$G) and tackle the challenge of constrained spectrum availability, it becomes imperative to explore effective spectrum access approaches.
On the other hand, the spectrum access rights are conferred through the issuance of dedicated licenses, primarily to entities such as mobile network operators (MNOs) \cite{khadem2023dynamic}. The scarcity of high-cost spectrum licenses, coupled with associated obligations in spectrum bands allocated for mobile broadband services, presents significant barriers for new entrants to this market, known as vertical markets \cite{4287565}.

With the introduction of $5$th generation network ($5$G) and the deployment of localized small cells, new opportunities are arising, especially for vertical sector players (VSPs) as micro-operators. In this context, each VSP operates in a private network to deliver its services directly to specific end-users. They also offer specialized services with a focus on high-quality service delivery. Furthermore, various vertical sectors such as audio program making and special events (PMSE), e-health, wireless industrial automation, public protection and disaster relief (PPDR), intelligent transport systems, car test tracks, drone control, and payload management are defined which exhibit distinct characteristics in terms of spectrum allocation and utilization \cite{ETSI_TR103885}.

Typically, resource management is widely used to maximize system functionality. In \cite{8377308}, a centralized power allocation scheme is presented to maximize system throughput while considering the quality of service (QoS) of users requirements in both uplink and downlink for an inband full-duplex (IBFD) multi-tier network. In \cite{huang2004adaptive}, the examination of adaptive resource allocation for multimedia QoS support in broadband wireless networks is undertaken. In \cite{6945296}, cloud users' requests are processed and transmitted by a cloud provider, all while adhering to specified QoS requirements.

In \cite{vuojala2020spectrum}, five optional spectrum access scenarios for 5G vertical network service providers (VNSPs) in acquiring spectrum access to deliver network services to either end-users or other enterprises have been defined, where the stakeholders engaged in the spectrum awarding process have been identified, and their roles and objectives have been examined. Moreover, the advantages and disadvantages of each scenario have been encompassed.
In \cite{8610409}, an analysis is conducted on the prevailing spectrum valuation methodologies, and the critical factors to be taken into account are identified when defining and evaluating the value of spectrum, particularly within the framework of forthcoming local 5G networks.
In \cite{matinmikko2018regulations}, the study investigates the impact of 5G innovations on mobile communications and examines regulatory aspects related to 5G development for locally deployed networks. It also extends the recent micro licensing model to local spectrum authorization in future 5G systems and provides guidelines for its essential components. This approach, known as the local micro licensing model, aims to open the mobile market by allowing diverse stakeholders to create local small cell networks using locally issued spectrum licenses. This ensures QoS is tailored to specific vertical sector needs.
In \cite{matinmikko2017micro}, the concept of micro-operators (uO) is introduced for 5G, focusing on local service delivery and indoor small cell communication. Key elements include regulation-related factors (local spectrum access rights) and technology-related factors (flexible network implementation). This concept offers business opportunities, such as providing local connectivity to all MNOs, operating secure networks for vertical sectors, and offering locally tailored content and services.

As can be seen, 
most existing studies introduce VSPs as micro-operators conceptually, with no consideration for spectrum management in the network. In this paper, we follow recommendations  suggested by standard bodies such as  \cite{ETSI_TR103885}, aimed at promoting opportunities for VSPs in 5G and 6G networks. We propose a novel priority-aware auction-based spectrum management for such networks. We aim to maximize data rate of the entire network by considering the minimum guaranteed QoS of each VSP, which supports the guaranteed bit rate (GBR) represented in 3GPP standards \cite{ts23.501}. To the best of our knowledge, this matter has not been examined in preceding studies.
As we highly depend on using deep learning techniques, 6G will indeed be a better framework to accommodate the proposed framework but the system model can be used for any 5G and beyond framework.

In contrast
to \cite{8377308, huang2004adaptive, 6945296}, we use a new compatible auction mechanism as a means to give precedence to QoS.
Moreover, compared to the scheme proposed in prior studies \cite{vuojala2020spectrum, 8610409, matinmikko2018regulations, matinmikko2017micro}, we deploy a novel and pragmatic system model, wherein the VSPs can be assumed as the elements within a licensed shared access (LSA) framework, and in an auction process, they can get their required data rate guaranteed as high-quality wireless networks.
Simulation results demonstrate the efficiency of the suggested priority-based spectrum management scheme when used within the proposed system model.

The rest of this paper is organized as follows. The system model and problem formulation for maximization of data rate is stated in Section \ref{section2}. In Section \ref{solution}, the proposed deep deterministic policy gradient (DDPG) algorithm for spectrum management is described. The simulation results of our resource management schemes is presented in Section \ref{results}. Finally, we conclude this paper in Section \ref{conclusion}.

\section{System Model} \label{section2}

\subsection{The Setup}
\begin{figure*}[t]
	\centering
	\includegraphics[width=0.77\linewidth]{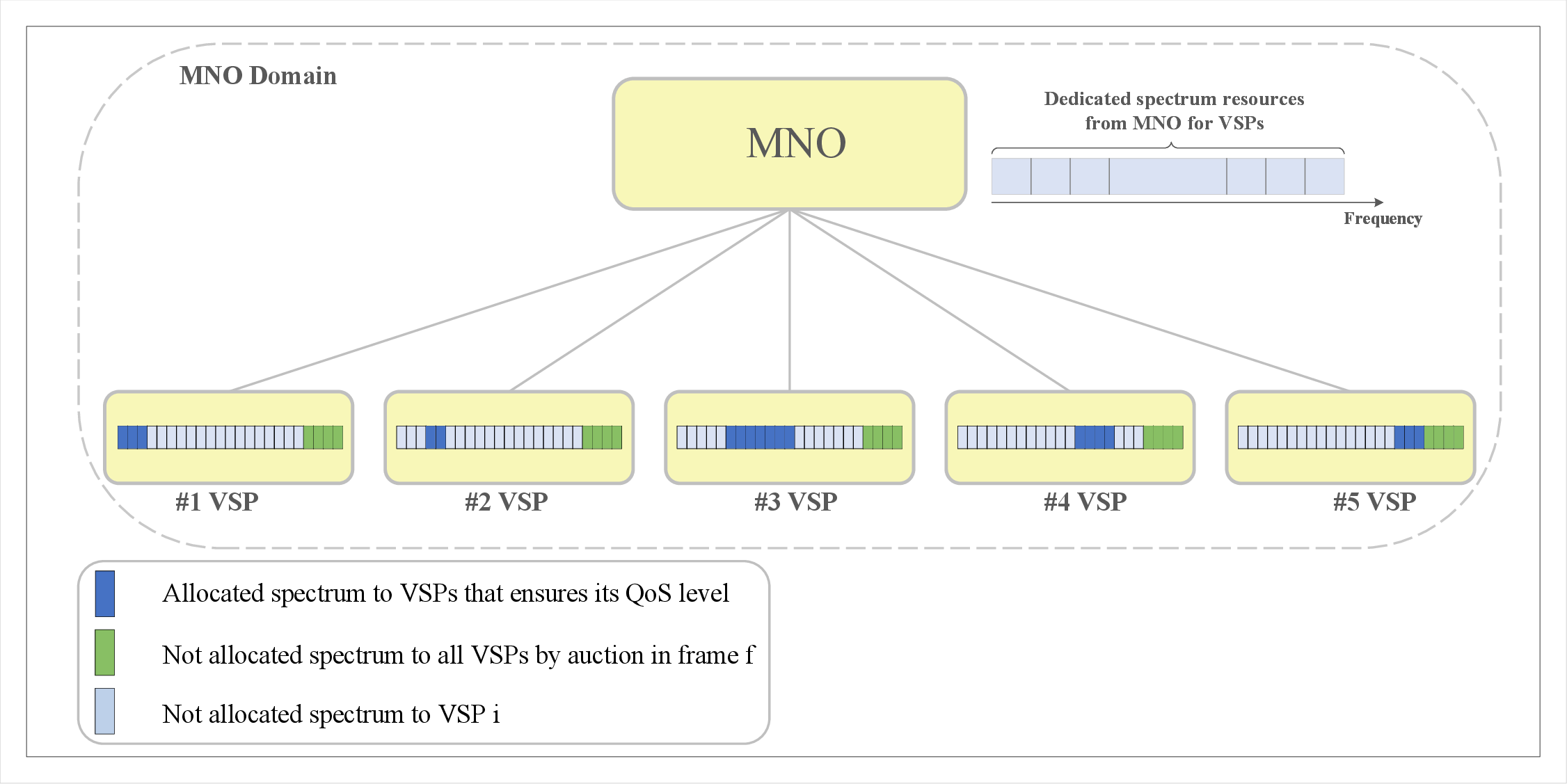}
	\caption{Integrating local high-quality networks (VSPs) into the MNO ecosystem in 6G network.}
	\label{system model}
\end{figure*}

The system model depicted in Fig. \ref{system model} includes an MNO that grants VSPs access to spectrum already licensed to it, which is not being used or planned for use in a particular area. We define the set $\mathcal{I} = \{1, \ldots, i, \ldots, I\}$ of VSPs that fall within the range of the MNO's operational area with different QoS characteristics defined as QoS Identifier (QI). Furthermore, the minimum data rates for VSPs in each frame are defined as a vector $\mathbf{r}^{\min} = (r_1^{\min}, \ldots, r_i^{\min}, \ldots, r_I^{\min})$, as represented in Table \ref{tab:GBR}. The maximum dedicated bandwidth of MNO for sharing with VSPs is a constant value of $B$, which is divided into several equal spectrum blocks with a constant bandwidth of $w$. The available bandwidth of MNO in each frame is denoted by ${BW}_f$. Furthermore, the number of available spectrum blocks in frame $f$ is equal to $\mathcal{J}_f = \frac{{BW}_f}{w}$.
 
We assume that spectrum management is done with the help of auction mechanism in a frame, and the results of auctions depend on guaranteed QoS of participated VSPs, available bandwidth of MNO, and received bids of VSPs in the frame we are in. 
\begin{table}[htbp]
	\centering
	\caption{Standardized 5QI of VSPs' mapping for GBR resource type\cite{ts23.501}}
	\scalebox{0.82}{
	\begin{tabular}{|c|c|c|c|c|}
		\hline
		\textbf{QCI} & \textbf{Packet Delay} & \textbf{Packet Error Loss Rate} & \textbf{Bit Rate} & \textbf{Priority} \\
		\hline
		VSP 1, 5 & 100 (ms) & $10^{-2}$ & 66.44 (pbs) & 1 \\
		\hline
		VSP 2 & 150 (ms) & $10^{-3}$ & 66.44 (pbs) & 3 \\
		\hline
		VSP 3 & 50 (ms) & $10^{-3}$ & 199.32 (pbs) & 2 \\
		\hline
		VSP 4 & 300 (ms) & $10^{-6}$ & 66.44 (pbs) & 4 \\
		\hline
	\end{tabular}
		}	
	\label{tab:GBR}
\end{table}
\subsection{Priority-based Auction for Spectrum Management}
In this section, we describe a multi-unit repeating Vickrey-Clarke-Groves (VCG) auction model in which the VSPs bid for one or more spectrum blocks within the available spectrum band. In our auction model, the spectrum broker is the MNO, who is the owner of the license to access the desired spectrum, while the VSPs are the bidders. Moreover, with assistance of the VCG auction mechanism, we can select multiple winners while the maximum number of available spectrum blocks are selected. We assume that the MNO receives requests from service providers in the form of a tuple $(b_{i,f}, p_{i,f}, n_{i,f}, r_{i,f})$, where the elements of the tuple represent the bid, value, demand (represented as the number of required spectrum blocks), and the obtainable data rate after winning in the auction, respectively. Afterwards, MNO applies a coefficient $c_{i,f}$ to bids for modifying the priority of VSPs so as to be able to influence the auction results, aiming to guarantee the minimum QoS of the winner VSPs, with considering their truthfulness as follows:
\begin{align}
	c_{i,f} = f(q_i, \theta_{i,f}),
	\label{coeficient}
\end{align}

where $q_{i}$ represents the constant QoS class identifier (QCI) priority for VSP $i$ based on the QI values in Table \ref{tab:GBR}. Moreover, we define $\theta_{i,f} = \frac{b_{i,f}}{v_{i,f}}$ to represent the truthfulness of the received bid of VSP $i$ by checking how close the bid is to the value. In fact, closer values of $\theta$ to 1 implies that VSP $i$ has more truthfulness and urgency in its request. Therefore, $c_{i,f}$ is a function of $q_i$ and $\theta_{i,f}$, which varies in every auction and is determined by the MNO to maximize the objective function. We have $c_{i,f} = \theta_{i,f} = 0$ if the VSP $i$ does not participate in the auction.
The received requests are collected in MNO as the auction broker and if the available spectrum, denoted as $BW_f$, surpasses the total requested spectrum, then the requested spectrum blocks assignments to the VSPs. Otherwise, the auction initiates when
\begin{align}
	w.\sum_{i \in \mathcal{I}}n_{i,f} > BW_f.
\end{align} 
Moreover, the winner determination process is formulated as a knapsack problem by considering the possible combinations of requests. Therefore, the best combination determines the winners of the auction, so we define $x_{i,f}$ as a binary winning indicator where $x_{i,f} = 1$ if the $i$-th VSP wins the auction in frame $f$, and otherwise, $x_{i,f} = 0$.
\subsection{Utility Function}
Each VSP has a minimum data rate requirement ($r_{i}^{\min}$) based on its service type, which includes considerations for packet delay and packet error loss rate, which determine its GBR priority. Therefore, each VSP is influenced by a fixed coefficient specific to it, presented in Table \ref{tab:GBR}. Furthermore, there is an achievable data rate for each VSP if the requested spectrum blocks are successfully allocated to it through an auction, which is a variable coefficient defined in the formula (\ref{coeficient}).
The purpose of optimization is to determine the appropriate variable coefficients for each VSP aimed to maximize the total data rate, while balancing the diverse requirements of VSPs.  

Therefore, we examine the utility function with a focus on proportional fairness \cite{6785480}, as the follows: 
\begin{align}
	\mathbb{U}_f = \sum_{i \in \mathcal{I}}(x_{i,f}r_{i,f} - r_{i}^{\min}).
	\label{utility}
\end{align}

\subsection{The Optimization Problem}
Based on previous definitions, our aim is to solve the following optimization problem
\begin{subequations}
	\begin{align}{\label{optimization}}
		&\max_{c} \mathbb{U}_f \\
		\mathbf{s.t.} \indent  &\mathbf{r}_{i,f} \geq \mathbf{r}_i^{\min}, \forall i \in \mathcal{I}, f > 0\\
		& 0 \leq \theta_{i,f} \leq 1, \forall i \in \mathcal{I}, f > 0\\
        & c_{i,f} \geq 0, \forall i \in \mathcal{I}, f > 0\\
        & x_{i,f} \in \{0 , 1\}, \forall i \in \mathcal{I}, f > 0,
	\end{align} 
\end{subequations}
where $f$ represents the frame in which we are, when maximizing the utility function (\ref{utility}).
\section{Solution} \label{solution}
In this section, we employ the actor-critic deep reinforcement learning (DRL) approach to construct a model for problem optimization (\ref{optimization}).
The actor-critic algorithm fuses Q-learning and policy gradient, two well-established reinforcement learning algorithms. This algorithm exhibits exceptional effectiveness when tackling intricate ML tasks \cite{10013255}. DDPG is a type of actor-critic DRL that learns a Q-function and a policy simultaneously. The actor makes choices regarding which action to perform, while the critic provides feedback to evaluate the actor's action quality. The DDPG framework consists of three fundamental components: the state space, the action space, and the instantaneous reward.

\begin{table}[hb]
	\centering
	\caption{Simulation Parameters}
	\label{simulation parameters}
	\scalebox{1}{
		\begin{tabular}{ll}
			\hline\hline
			\multicolumn{1}{l}{\textit{\textbf{Environment parameters}}} & \multicolumn{1}{l}{\textit{\textbf{Value}}}  \\  \hline \vspace{-0.2 cm}\\
			Number of VSPs          &5 \\    
			Total bandwidth &300 MHz \\
			Bandwidth of blocks           & 5 MHz \cite{4287565} \\
			Maximum rate of each VSP        &500 bps \\
			\hline \vspace{-0.2 cm}\\
			\multicolumn{1}{l}{\textit{\textbf{DNN parameters}}} & \multicolumn{1}{l}{\textit{\textbf{Value}}}  \\  \hline \vspace{-0.2 cm}\\
			Experience replay buffer size   &10000 \\
			Mini batch size                 &64 \\ 
			Number/size of actor networks hidden layers &2 / 1024, 512 \\
			Number/size of critic networks hidden layers &2 / 512, 256 \\
			DQN networks learning rate &0.001 \\
			Epsilon decreasing rate     &0.0005 \\
			Minimum epsilon rate      &0.01 \\
			Critic/Actor networks learning rate  &0.001/0.0001 \\
			Target networks update parameter &0.0005 \\
			Discount factor                 &0.99 \\
			Number of episodes               &500 \\ 
			Number of frames per episode      &250 \\
			\hline\hline	
		\end{tabular}
	}
\end{table}
\indent $\bullet$ \textbf{State space:} To enhance the process of choosing auction winners based on their condition, VSPs request are defined as state space. $S_f$ can be represented as a state space in frame $f$ by a vector in the form 
\begin{align}
	s_f = \{s_{1,f},s_{2,f},...,s_{i,f},...,s_{I,f}\},
\end{align}
where $s_{i,f}$ indicates the $i$-th VSP request in frame $f$ consisting of bid, auction value and its demand as mentioned in Section \ref{section2}. 

\indent $\bullet$\textbf{ Action space:} For VSPs request, we define coefficient for each VSP in frame $f$ as action space by a vector in the form 
\begin{align}
	a_f = \{c_{1,f},c_{2,f},...,c_{i,f},...,c_{I,f}\},
\end{align}
where $a_{i,f}$ indicates the $i$-th VSP coefficient in frame $f$. \\
\indent $\bullet$\textbf{ Instantaneous reward:} The Instantaneous reward reflects the correct running state and the improvement of choosing winners. $R_f$ is defined as an instantaneous reward in frame $f$ according to the utility function
\begin{align}
	R_f = \sum_{i \in \mathcal{I}}(x_{i,f}r_{i,f} - r_{i}^{\min}),
\end{align}
where $x_{i,f}$ is the winning indicator for $i$-th VSP and $r_{i,f}$ and $r_{i}^{\min}$ are data rate and minimum data rate for $i$-th VSP in frame $f$, respectively.

The actor network's decision-making process relies on the following equation:
\begin{align}
	a_f = \pi(s_f,\psi).
\end{align}
Here, $\pi$ denotes the policy of the actor network, and $\psi$ signifies the actor network's weight. At each frame, the agent (MNO) takes an action $a_f$, obtains an instantaneous reward $R_f$, and transitions to a new state $s_{f+1}$. The tuple $(s_f,a_f,R_f,s_{f+1})$ is then stored in the replay memory buffer $D$ with a capacity of $N$. The agent selects a random sample $i$ from the buffer, and the actor network can learn from the following loss function:
Here, $Q$ represents the estimated value provided by the critic network with weights $\varphi$, which evaluates the actor network's policy
\begin{align}
	\label{loss1}
	L^{A}_{\psi} = -(Q(s_f,\pi(s_f,\psi),\varphi)).
\end{align}
The agent's objective is to either maximize $Q$ or minimize $-Q$ in order to adjust the actor network's weights through the gradient update equation:
\begin{align}
	\psi \leftarrow \psi - \tau_1 \nabla_{\psi} L_{\psi}^{\text{A}}.
\end{align}
For the critic network, learning is accomplished using the following loss function:
\begin{align}
	\label{loss2}
	L^{C}_{\varphi} = \sum_f ((r_f+\gamma \tilde{Q}(s_{f+1},\tilde{\pi}(s_{f+1},\tilde{\psi})\tilde{\varphi})) - Q(s_f,a_f,\varphi))^2,
\end{align}
where $\tilde{Q}$ represents the Q-value from the target critic network, and $\tilde{\varphi}$ and $\tilde{\psi}$ are the weights of the target critic network and target actor network, respectively. The critic network's weights can be updated using the gradient as follows:
\begin{align}
	\varphi \leftarrow \varphi - \tau_2 \nabla_{\varphi} L_{\varphi}^{\text{C}}.
\end{align}
Ultimately, the actor and critic target network weights are adjusted as follows:
\begin{align}
	\label{update1}
	\tilde{\psi} \leftarrow \psi + (1-\tau_3) \tilde{\psi},
\end{align}
\begin{align}
	\label{update2}
	\tilde{\varphi} \leftarrow \varphi + (1-\tau_3) \tilde{\varphi}.
\end{align}

Here, $\tau_3$ determines the proportion of weights to transfer from the main networks to the target networks. In the subsequent frame, the algorithm generates fresh experiences and updates the network weights using new data from the batch. To sum it up, all the previously discussed steps can be summarized in the following algorithm:
\begin{algorithmframe}
\begin{algorithmic}[1]
	\label{DDPG}
	\STATE Initialize the weights $\psi$ of actor network and the weights $\varphi$ of critic network. \\
	\STATE Initialize the weights $\tilde{\psi}$ of target actor network and the weights $\tilde{\varphi}$ of target critic network, $\tilde{\psi} = \psi$, $\tilde{\varphi} = \varphi$. \\
	\FOR{each episode} \STATE Get VSPs request in first frame, $S_0$. \\
	\FOR{each frame $f$}
	\STATE Observe $s_f = \{s_{1,f},...,s_{i,f},...,s_{I,f}\},$ and choose action $a_f = \pi(s_f,\psi)$ according to current policy and exploration noise.\\
	\STATE Receive $s_{f+1}=\{s_{1,f+1},...,s_{i,f+1},...,s_{I,f+1}\}$ and \STATE Calculate reward $R_f$. \\
	\STATE Store transition $(s_f,a_f,R_f,s_{f+1})$ in replay buffer $D$ with size $N$.\\
	\STATE Sample minibatch of size $B$, from $D$. \\
	\STATE Update weights $\psi$ and $\varphi$ by minimizing the loss function in (\ref{loss1}), (\ref{loss2}).\\
	\STATE Update weights $\tilde{\psi}$ and $\tilde{\varphi}$ by (\ref{update1}), (\ref{update2}).
	\ENDFOR \ENDFOR 
\end{algorithmic}
\end{algorithmframe}

\section{Simulation Results} \label{results}
In this section, we demonstrate the performance of our DRL system model. Our algorithm is put into practice using Python version 3.7.9 and the Spyder Integrated Development Environment (IDE) version 5.2.2. The simulation parameters are represented in Table \ref{simulation parameters}.

\subsection{Winning Percentage of Each VSP‌ Based on Priority:}
Based on our system model, we assume that there are $5$ VSPs within the MNO network, and each VSP entity requires a distinct QoS, based on the nature of its providing service. When we examine the initial QCI assumption for each VSP, it becomes evident that the success data rate of each VSP is directly tied to its prioritization within Table \ref{tab:GBR}.
\begin{figure}[b]
	\centering
	\includegraphics[width=1\linewidth]{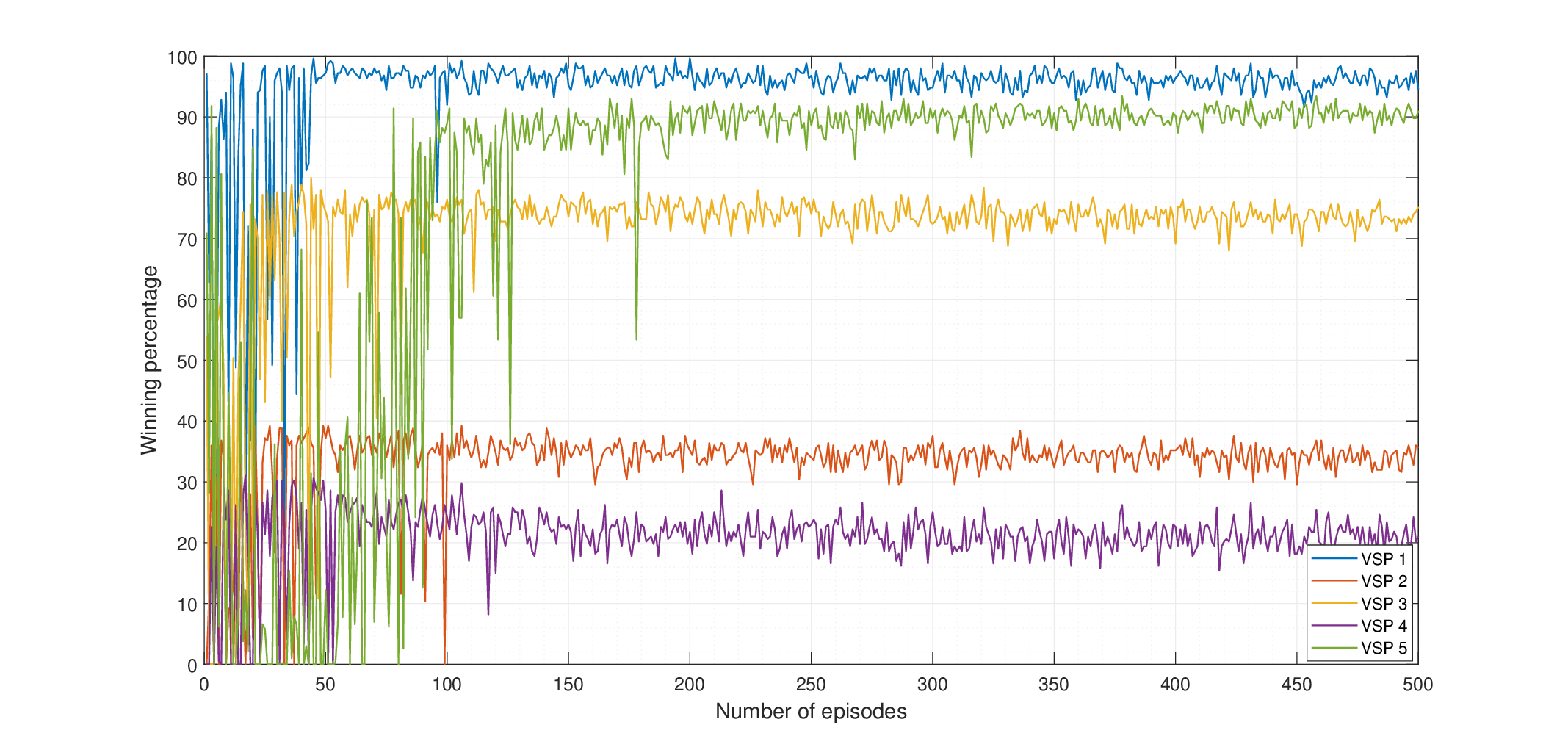}
	\caption{The winning percentage of each VSP during the auctions.}
	\label{F1} 
\end{figure} 
It can be seen in Fig. \ref{F1} that although VSP $1$ and VSP $5$ have the same QoS value, VSP $5$ exhibits lower truthfulness ($\theta$) during the auction process, resulting in a lower success rate than VSP $1$.

\subsection{Mean Reward by Considering Network Size:}
\begin{figure}[t]
	\centering
	\includegraphics[width=1\linewidth]{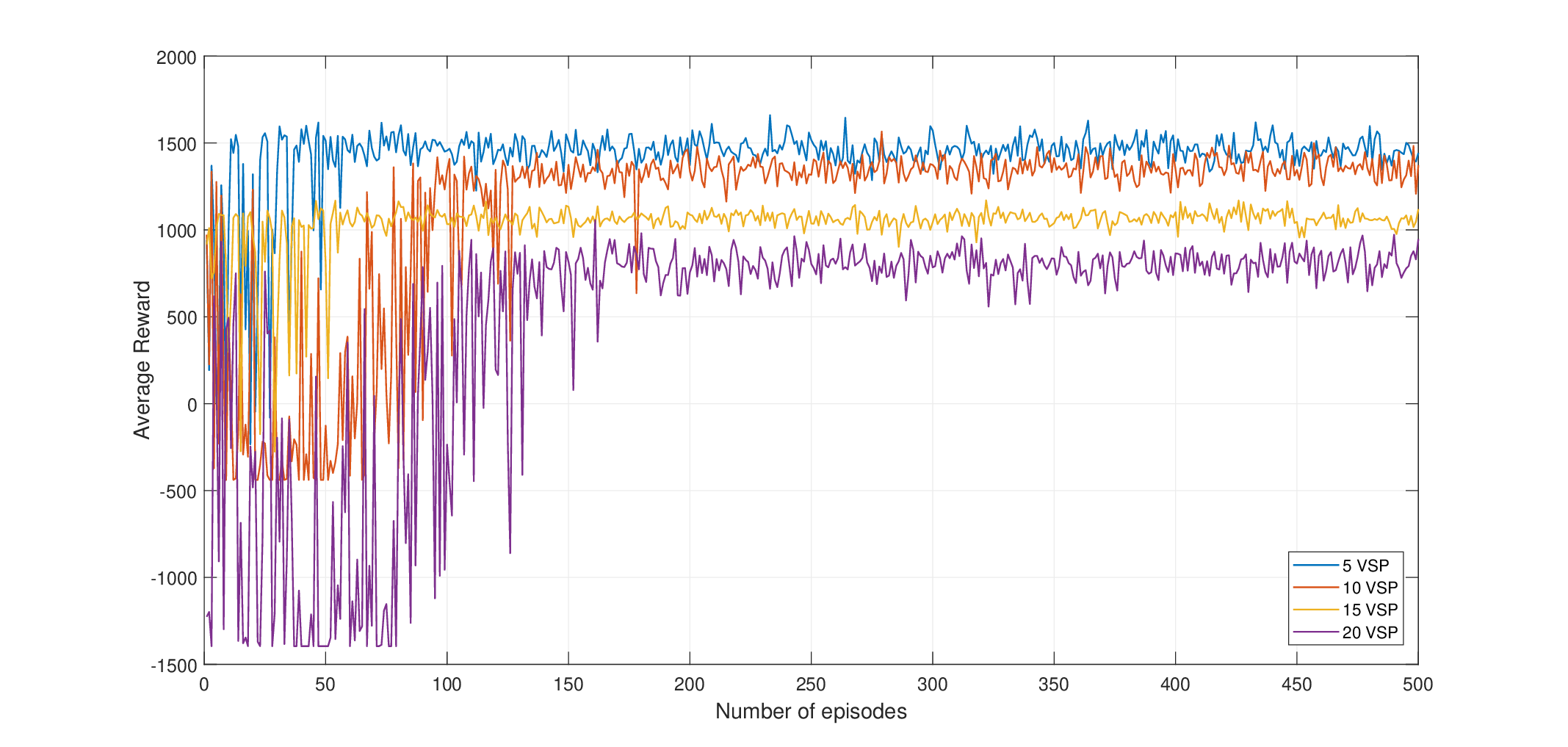}
	\caption{Mean reward per number of episodes.}
	\label{F2} 
\end{figure}
In Fig. \ref{F2}, we aim to assess the impact of network size expansion on convergence outcomes. Our simulation results indicate that as the network size grows and spectrum management complexities increase, convergence occurs later in the process. Nevertheless, the network attains favorable outcomes. Specifically, when we extend the number of VSPs to $20$, adhere to the minimum rate values specified in Table \ref{tab:GBR}, and set an arbitrary maximum rate (here $500$ bps) for each VSP, the network consistently exhibits favorable numerical performance for the utility function.

\subsection{Spectrum Utilization}
\begin{figure}[t]
	\centering
	\includegraphics[width=0.8\linewidth]{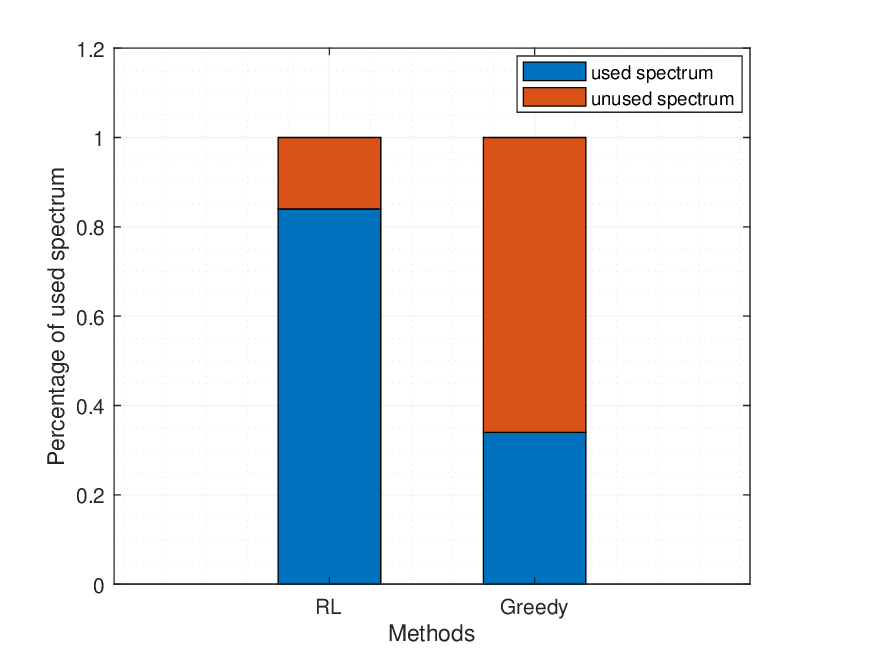}
	\caption{Spectrum efficiency for three methods.}
	\label{F5} 
\end{figure}
In this part, we examined the percentage of spectrum usage by considering the available spectrum blocks in 500 auctions. We used two methods, as shown in Fig \ref{F5}. In the first bar, we utilized the RL algorithm, i.e. DDPG, to determine the coefficients of bids based on the results of previous auctions. \textcolor{black}{In the 2nd bar, we used the greedy approach \cite{cary2007greedy} as a baseline framework, which assigns spectrum blocks without using RL algorithms.}

Consequently, the results show that the performance of the RL-based method is about $\%85$, which is a significant level of spectrum utilization. In contrast, the greedy algorithm yields a value of $\%35$, which is not favorable.

\subsection{The Effect of Increasing the Number of Spectrum Blocks on Fairness Index}
\begin{figure}[h]
	\centering
	\includegraphics[width=0.8\linewidth]{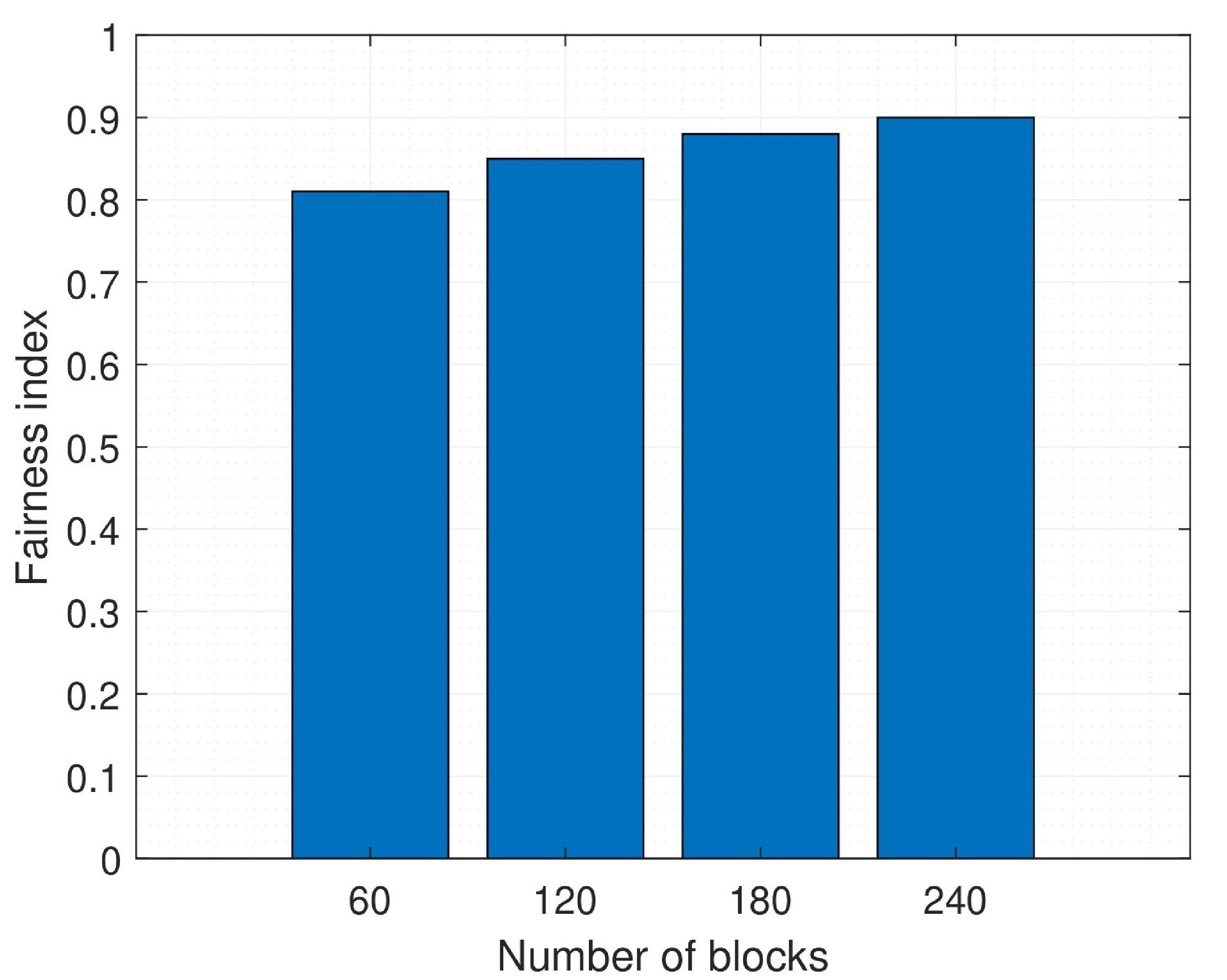}
	\caption{Fairness index per Number of VSPs.}
	\label{F4} 
\end{figure}
To demonstrate the impact of the number of spectrum blocks on the fairness index of the network, we vary it to $60$, $120$, $180$, and $240$ spectrum blocks. As shown in Fig. \ref{F4}, our proposed approach performs better with an increase in the number of spectrum blocks.
In particular, we achieve a fairness index of $\%90$, for 240 blocks, which is a significant value for network fairness.

\section{Conclusion}\label{conclusion}
In this paper, we investigated the problem of maximizing data rate of VSPs by satisfying their minimum data rate and allocating spectrum in 5G and 6G networks. We considered VSPs with different QoSs. To address the complexity
increase when we have higher number of VSPs and resource blocks,
we deployed the DRL methods such as the DDPG algorithm. In the simulation results, we examined various factors that affect our proposed network, including the role of QoS priority factors, truthfulness, and bidding in auction outcomes, as well as the impact of network complexity on algorithm convergence and fairness. 
The results showed that the number VSPs' wins is entirely depending on the priority of their services, as well as their level of truthfulness. Based on this, the fairness within the network reaches an appropriate level for the larger number of spectrum blocks. Furthermore, our proposed method enhances spectrum efficiency by approximately $\%70$ when compared to the greedy method.

\bibliographystyle{ieeetr}
\bibliography{Citation}
\end{document}